\documentclass[twocolumn, times, tighten]{aastex62}
\usepackage{amsmath}	
\usepackage{enumitem}

\usepackage[]{xcolor}

\definecolor{mygreen}{cmyk}{0.75002,0,1,0}

\newcommand{\ha}{H$\alpha$}

\newcommand{\hb}{H$\beta$}
\newcommand{\mgii}{Mg\,{\sc ii}}
\newcommand{\civ}{C\,{\sc iv}}

\newcommand{\oiiiwu}{[O{\,\sc iii}]$\lambda5007$}

\newcommand{\kmps}{$\rm km\,s^{-1}$}
\newcommand{\xmm}{{\it XMM-Newton}}
\newcommand{\chandra}{{\it Chandra}}
\newcommand{\rosat}{{\it ROSAT}}

\received{xxx}
\revised{xxx}
\accepted{xxx}
\submitjournal{ApJS}

\shorttitle{LAMOST DR4\&5}
\shortauthors{Su et al.}

\begin{document}

\title{The Large Sky Area Multi-Object Fiber Spectroscopic Telescope (LAMOST) Quasar Survey: the Forth and Fifth Data Release}

\correspondingauthor{Su Yao}
\email{yaosu@pku.edu.cn}

\author[0000-0002-9728-1552]{Su Yao}
\altaffiliation{KIAA-CAS fellow}
\affil{Kavli Institute for Astronomy and Astrophysics, 
Peking University, 
Beijing 100871, China}
\affiliation{National Astronomical Observatories, 
Chinese Academy of Sciences, 
Beijing 100012, China}

\author[0000-0002-7350-6913]{Xue-Bing Wu}
\affiliation{Kavli Institute for Astronomy and Astrophysics, 
Peking University, 
Beijing 100871, China}
\affiliation{Department of Astronomy, School of Physics, 
Peking University, 
Beijing 100871, China}

\author{Y. L. Ai}
\affiliation{School of Physics and Astronomy, 
Sun Yat-Sen University, 
Guangzhou 510275, China}

\author{Jinyi Yang}
\affiliation{Department of Astronomy, School of Physics, 
Peking University, 
Beijing 100871, China}
\affiliation{Kavli Institute for Astronomy and Astrophysics, 
Peking University, 
Beijing 100871, China}

\author[0000-0002-6893-3742]{Qian Yang}
\affiliation{Department of Astronomy, School of Physics, 
Peking University, 
Beijing 100871, China}
\affiliation{Kavli Institute for Astronomy and Astrophysics, 
Peking University, 
Beijing 100871, China}

\author{Xiaoyi Dong}
\affiliation{Department of Astronomy, School of Physics, 
Peking University, 
Beijing 100871, China}

\author{Ravi Joshi}
\affiliation{Kavli Institute for Astronomy and Astrophysics, 
Peking University, 
Beijing 100871, China}

\author[0000-0002-7633-431X]{Feige Wang}
\affiliation{Department of Astronomy, School of Physics, 
Peking University, 
Beijing 100871, China}
\affiliation{Kavli Institute for Astronomy and Astrophysics, 
Peking University, 
Beijing 100871, China}

\author{Xiaotong Feng}
\affiliation{Department of Astronomy, School of Physics, 
Peking University, 
Beijing 100871, China}
\affiliation{Kavli Institute for Astronomy and Astrophysics, 
Peking University, 
Beijing 100871, China}

\author{Yuming Fu}
\affiliation{Department of Astronomy, School of Physics, 
Peking University, 
Beijing 100871, China}
\affiliation{Kavli Institute for Astronomy and Astrophysics, 
Peking University, 
Beijing 100871, China}

\author{Wen Hou}
\affiliation{National Astronomical Observatories, 
Chinese Academy of Sciences, 
Beijing 100012, China}

\author{A.-L. Luo}
\affiliation{National Astronomical Observatories, 
Chinese Academy of Sciences, 
Beijing 100012, China}

\author{Xiao Kong}
\affiliation{National Astronomical Observatories, 
Chinese Academy of Sciences, 
Beijing 100012, China}

\author{Yuanqi Liu}
\affiliation{Department of Astronomy, School of Physics, 
Peking University, 
Beijing 100871, China}

\author{Y.-H. Zhao}
\affiliation{National Astronomical Observatories, 
Chinese Academy of Sciences, 
Beijing 100012, China}

\author{Y.-X. Zhang}
\affiliation{National Astronomical Observatories, 
Chinese Academy of Sciences, 
Beijing 100012, China}

\author{H.-L. Yuan}
\affiliation{National Astronomical Observatories, 
Chinese Academy of Sciences, 
Beijing 100012, China}

\author{Shiyin Shen}
\affiliation{Key Laboratory for Research in Galaxies and Cosmology, 
Shanghai Astronomical Observatory, \\
Chinese Academy of Sciences, 
Shanghai 200030, China}

\begin{abstract}

We present the Data Release 4\&5 quasar catalog from the quasar survey by Large Sky Area Multi-Object Fiber Spectroscopic Telescope (LAMOST), 
which includes quasars observed between September 2015 and June 2017. 
There are a total of {19,253} quasars identified by visual inspections of the spectra. 
Among them, 
{11,458} are independently discovered by LAMOST, 
in which {3296} were reported by SDSS DR12 and DR14 quasar catalog after our survey began, 
while the rest {8162} are new discoveries of LAMOST. 
We provide the emission line measurements for the \ha, \hb, \mgii\ and/or \civ\ for 18100 quasars. 
Since LAMOST does not have absolute flux calibration information, 
we obtain the monochromatic continuum luminosities by fitting the SDSS photometric data using the quasar spectra, 
and then estimate the black hole masses. 
The catalog and spectra for these quasars are available online.
This is the third installment in the series of LAMOST quasar survey 
which has released spectra for totally {$\sim43,000$} quasars hitherto. 
There are {24,772} independently discovered quasars, 
{17,128} of which are newly discovered. 
In addition to the great supplement to the new quasar discoveries, 
LAMOST has also provided a large database (overlapped with SDSS) for investigating the quasar spectral variability and discovering unusual quasars, including changing-look quasars, 
with ongoing and upcoming large surveys.

\end{abstract}




\section{Introduction} \label{sec:intro}

Quasars, 
as powered by the accretion onto the supermassive black holes (SMBHs) residing in the center of their host galaxies, 
can emit radiation in a broad range of wavelength from radio to $\gamma$-rays, 
and are the most luminous, long-lived celestial objects in the Universe. 
They are key ingredients in our understanding of 
the evolution of the galaxies through cosmic time \citep[][]{2013ARA&A..51..511K}, 
and can be used to probe the intergalactic medium, 
the large scale structure and the reionization history of the early universe \citep[][]{2001AJ....122.2850B,2002AJ....123.1247F, 2006AJ....132..117F, 2009ApJ...697.1634R}. 

Since the discovery of the first quasar \citep{1963Natur.197.1040S}, 
great efforts have been made to find larger number of quasars during the past half century 
\citep[e.g.,][]{1983ApJ...269..352S, 
1995AJ....109.1498H, 
1996ApJ...468..121S, 
2000MNRAS.317.1014B, 
2000AJ....120.1579Y}. 
Especially, 
the milestone of discovering more than $10^{5}$ new quasars has been reached 
thanks to the advent of two large surveys in last two decades, namely, 
the Two-Degree Field Quasar Redshift Survey \citep[2dF,][]{2004MNRAS.349.1397C} 
and 
the Sloan Digital Sky Survey \citep[SDSS,][]{2010AJ....139.2360S}. 
With the latest data release of SDSS-IV, 
the total number of currently known quasars reaches more than half million \citep[][]{2018A&A...613A..51P}.

Quasars have distinct colors from most of stars and normal galaxies \citep[][]{1999AJ....117.2528F}. 
So the quasar candidates for spectroscopic observations are usually selected based on their multi-color photometric data. 
In particular, 
quasars at $z<2.2$ have a UV excess that distinguishes them from most stars. 
For instance, 
both 2dF and SDSS-I/II select quasar candidates mainly based on UV/optical photometric data in the color-color diagram 
\citep[][]{2002AJ....123.2945R, 2005MNRAS.359...57S}. 
However, 
the completeness and efficiency of this method becomes low at $2.2<z<3.5$, 
especially at $z=2.7$ in SDSS, 
as the stellar locus intersects with the region occupied by quasars in the color-color diagram 
\citep[][]{2007AJ....134..102S}.

To maximize the efficiency of identifying quasars at $2.2<z<3.5$, a lot of improvements have been made. 
Analogous to the UV excess seen in low-redshift quasars, 
there is also an excess in the near infrared $K$-band for quasars at $z>2.2$ compared to stars. 
A method of selecting quasar candidates at $z>2.2$ using $K$-band excess based on the UKIDSS was proposed \citep[e.g.,][]{2000MNRAS.312..827W, 2002MNRAS.337.1153S, 2007MNRAS.379.1599L, 2008MNRAS.386.1605M, 2008MNRAS.389..407S}. 
By combining the optical and infrared photometric data, 
\citet{2010MNRAS.406.1583W} and \citet{2012AJ....144...49W} proposed effective criteria for selecting quasar candidates based on SDSS/UKIDSS and SDSS/$WISE$ colors. 
These methods have led to the discovery of more and more intermediate- and high-redshift quasars 
\citep[e.g.,][]{2011Natur.474..616M, 2013AJ....146..100W, 2015Natur.518..512W, 2016ApJ...819...24W}.

In recent years, 
the quasar candidates have also been searched by using various data-mining algorithms, e.g., 
extreme deconvolution method \cite[XDQSO,][]{2011ApJ...729..141B}, 
neural network combinator \citep[][]{2010A&A...523A..14Y} 
and 
Kernel Density Estimator \citep[KDE,][]{2004ApJS..155..257R, 2009ApJS..180...67R}, 
which are adopted by SDSS-III/BOSS \citep[][]{2012ApJS..199....3R}. 
Recently, \citet{2012MNRAS.425.2599P} develop a classification system that is made up of several support vector machine (SVM) classifiers, 
which can be applied to search for quasar candidates from large sky surveys. 
In addition, the variability-based selection \citep[][]{2012ApJS..199....3R} and cross-matching with X-ray and radio data \citep[][]{2016AJ....151...24A} are also performed as supplement for identifying quasars.

This paper presents the quasar catalog from Large Sky Area Multi-Object Fiber Spectroscopic Telescope (LAMOST) quasar survey conducted between September 2015 and June 2017. 
It is the third part in a series of LAMOST quasar survey paper, 
after the data release 1 \citep[DR1,][hereafter Paper I]{2016AJ....151...24A}, 
data release 2 and 3 \citep[DR2 and DR3,][hereafter Paper II]{2018AJ....155..189D}. 
In Section~\ref{sec:survey}, we briefly review the candidate selections and the spectroscopic survey. 
The emission line measurements and black hole mass estimations are described in Section~\ref{sec:fitting} and Section~\ref{sec:bhmass}. 
Section~\ref{sec:catalog} presents the description of the catalog and the parameters. 
In Section~\ref{sec:summary} we give the summary and discussions. 
Throughout this work a $\Lambda$-dominated cosmology is assumed with $H_0=70$ km s$^{-1}$ Mpc$^{-1}$, $\Omega_\Lambda=0.7$ and $\Omega_{\rm M}=0.3$.

\section{Survey Outline} \label{sec:survey}

LAMOST is a quasi-meridian reflecting Schmidt telescope with an effective light-collecting aperture that varies from 3.6\,m to 4.9\,m (depending on the pointing direction) and a $5\arcdeg$ field of view in diameter 
\citep[][]{1996ApOpt..35.5155W, 2004ChJAA...4....1S, 2012RAA....12.1197C}. 
It has $4000$ robotic fibers, with $\sim3\arcsec$ diameter, mounted on its focal plane and connected to 16 spectrographs. 
Each spectrum of the target is split into two channels, red and blue, 
and then recorded on red and blue cameras, respectively. 
The blue channel is optimized for $3700-5900$\,\AA, 
and the red channel for $5700-9000$\,\AA, with $200$\,\AA~overlaps between the two channels. 
The spectral resolution reaches $R\sim1800$ over the entire wavelength range 
\citep[see][]{2012RAA....12.1197C}. 

After commissioning from 2009 to 2010 and the pilot survey in 2011 
\citep[][]{2012RAA....12.1243L}, 
the LAMOST regular survey was carried out from September 2012 to June 2017, 
which was designed to have two major components: 
the LAMOST Experiment for Galactic Understanding and Exploration (LEGUE) 
and the LAMOST ExtraGAlactic Survey (LEGAS) \citep[][]{2012RAA....12..723Z}. 
The data is released in yearly increments. 
The LAMOST quasar survey was conducted under LEGAS which covers the high Galactic latitude area in the northern sky. 
The quasar candidates were observed simultaneously with other type of objects from the LEGUE and LEGAS samples during the survey. 
The typical exposure time for each target is usually $\sim90$ minutes which is equally divided into three exposures. 
The total exposure time would be adjusted according to the observation conditions, e.g. seeing. 
Although LEGAS used only a small fraction of the total observing time due to the weather condition and bright sky background, 
LAMOST has still collected useful data and identified more than 20,000 quasars, 
half of which are new discoveries, 
during the first three years. 
The results of the LAMOST quasar survey as released in DR1, DR2 and DR3 are presented in Paper I and II. 
In this paper, 
we present the results of the LAMOST quasar survey conducted in the forth and fifth year 
as released in LAMOST data release 4 (DR4) and data release 5 (DR5).

\subsection{Target Selection} \label{selection}

The details of the method used to select the targets for LAMOST quasar survey can be found in \citet{2010MNRAS.406.1583W}, \citet{2012AJ....144...49W}, \citet{2012MNRAS.425.2599P} and Paper I. 
Here we briefly review the target selection as follows. 

Most of the quasar candidates for spectroscopic followup are selected based on the photometry data of SDSS \citep[e.g.,][]{2012ApJS..203...21A}. 
Firstly, the targets are required to be point sources in SDSS images to avoid large numbers of galaxies. 
A faint limit of $i=20$ is adopted to avoid the too low signal-to-noise ratio, 
and a bright limit of $i=16$ is adopted to avoid the saturation and the cross-talk of the neighboring fibers. 
Then, after the correction for Galactic extinction \citep[][]{1998ApJ...500..525S}, 
the SDSS point-spread function (PSF) magnitudes are used in the selection algorithm. 
The targets are selected mainly based on the following methods: 
\begin{enumerate}
    \item {\it Optical-infrared colors.} 
    Although the quasars in redshift of $2.2<z<3.5$ have similar optical color as normal stars, 
    they are usually more luminous in infrared band \citep[e.g.,][]{2000MNRAS.312..827W, 2008MNRAS.386.1605M}. 
    Thus the distinct optical-infrared colors could be an useful tool in selecting quasar candidates. 
    The photometric point sources are cross-matched with the dataset from UKIDSS/Large Area Survey \citep[][]{2007MNRAS.379.1599L} and {\it WISE} all-sky data release\footnote{For the {\it WISE} data, only sources flagged as unaffected by artifacts in all four bands are used. } 
    \citep[][]{2010AJ....140.1868W}. 
    Then, the selection is made according to the location of sources in the multi-dimensional SDSS and UKIDSS/{\it{WISE}} color space \citep[][]{2010MNRAS.406.1583W, 2012AJ....144...49W}. 
    For the UKIDSS-matched sources, 
    quasar candidates are selected with criteria of 
    $Y-K>0.46(g-z)+0.82$ or $J-K>0.45(i-Y-0.366)+0.64$ \citep[][]{2010MNRAS.406.1583W}, 
    where $Y$, $J$, $K$ are in Vega magnitudes and $g$, $i$, $z$ are in AB magnitudes. 
    For the {\it WISE}-matched sources, 
    the candidates are selected with $w1-w2>0.57$ or $z-w1>0.66(g-z)+2.01$ \citep[][]{2012AJ....144...49W}, 
    where $w1$ and $w2$ are in Vega magnitudes and $g$, $z$ are in AB magnitudes. 
    \item {\it Date-mining algorithms.} 
    A mixture of heterogeneous data-mining algorithm, 
    such as support vector machine 
    \citep[SVM,][]{2012MNRAS.425.2599P}, 
    extreme deconvolution \citep[XD,][]{2011ApJ...729..141B} and 
    kernel-density-estimation technique \citep[KDE,][]{2004ApJS..155..257R, 2009ApJS..180...67R}, 
    are also adopted in combination with the optical-infrared color selection method. 
    We use the same SVM algorithm demonstrated in \citet{2012MNRAS.425.2599P}. 
    The XDQSO sample are identical to that provided by \citet{2011ApJ...729..141B}. 
    \item {\it Multi-wavelength data matching.} 
    In addition to the methods mentioned above, 
    the final input catalog of the targets for the LAMOST quasar survey are also supplemented by cross-matching the SDSS photometry with sources detected in X-ray surveys 
    ({\it XMM-Newton}, {\it Chandra}, {\it ROSAT}) 
    and radio survey (FIRST, NVSS). 
    The matching radius is 3\arcsec\ for FIRST, NVSS, \xmm\ and \chandra, and 30\arcsec\ for \rosat.
\end{enumerate}

\begin{figure}
	\centering
	\includegraphics[width=\columnwidth]{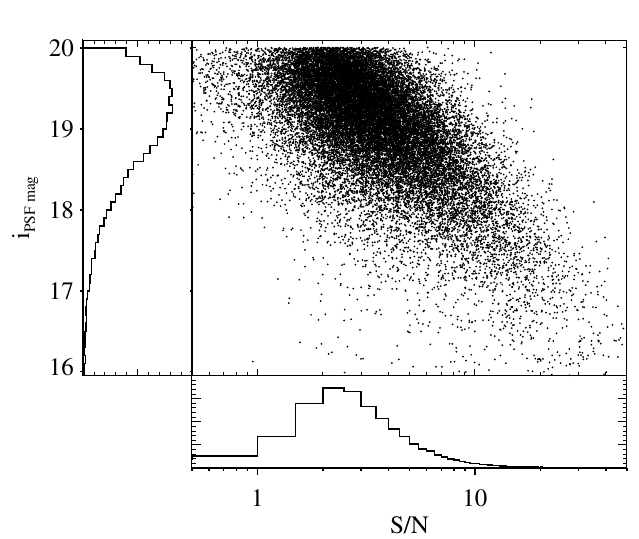}
	\caption{
	Distributions of the SDSS $i$-band PSF magnitude and the spectrum median signal-to-noise ratio (S/N) for the observed quasar candidates in forth and fifth years of LAMOST regular survey. }
	\label{fig:snr}
\end{figure}

In Figure~\ref{fig:snr}, 
we present the distribution of the $i$-band magnitude and the spectrum signal-to-noise ratio (S/N) for the observed quasar candidates in the forth and fifth years of LAMOST regular survey. 
Here, the S/N is calculated as the median S/N per pixel in the wavelength regions of $4000-5700$\,\AA\ and $6200-9000$\,\AA. 
As can be seen, 
the distributions of apparent $i$-band magnitude and spectrum S/N peak at $\sim19.4$\,mag and $\sim2.5$, respectively.

\subsection{Quasar Identification}

After the observations, 
the raw data were reduced by the LAMOST 2D pipeline. 
The final spectrum of each target is produced after procedures including 
dark and bias subtraction, flat field correction, spectral extraction, sky subtraction, wavelength calibration, merging sub-exposures and combining blue and red spectra 
(see \citealt{2015RAA....15.1095L} for details). 
Next, the spectra were passed to the LAMOST 1D pipeline, 
which automatically classifies the spectra into four categories according to their object types, namely `STAR', `GALAXY', `QSO' and `Unknown', 
and measures the redshift 
if the spectrum is classified into `GALAXY' or `QSO' \citep[][]{2015RAA....15.1095L}. 

\begin{figure*}[th]
\centering
	\begin{tabular}{cc}
		\includegraphics[width=0.46\textwidth]{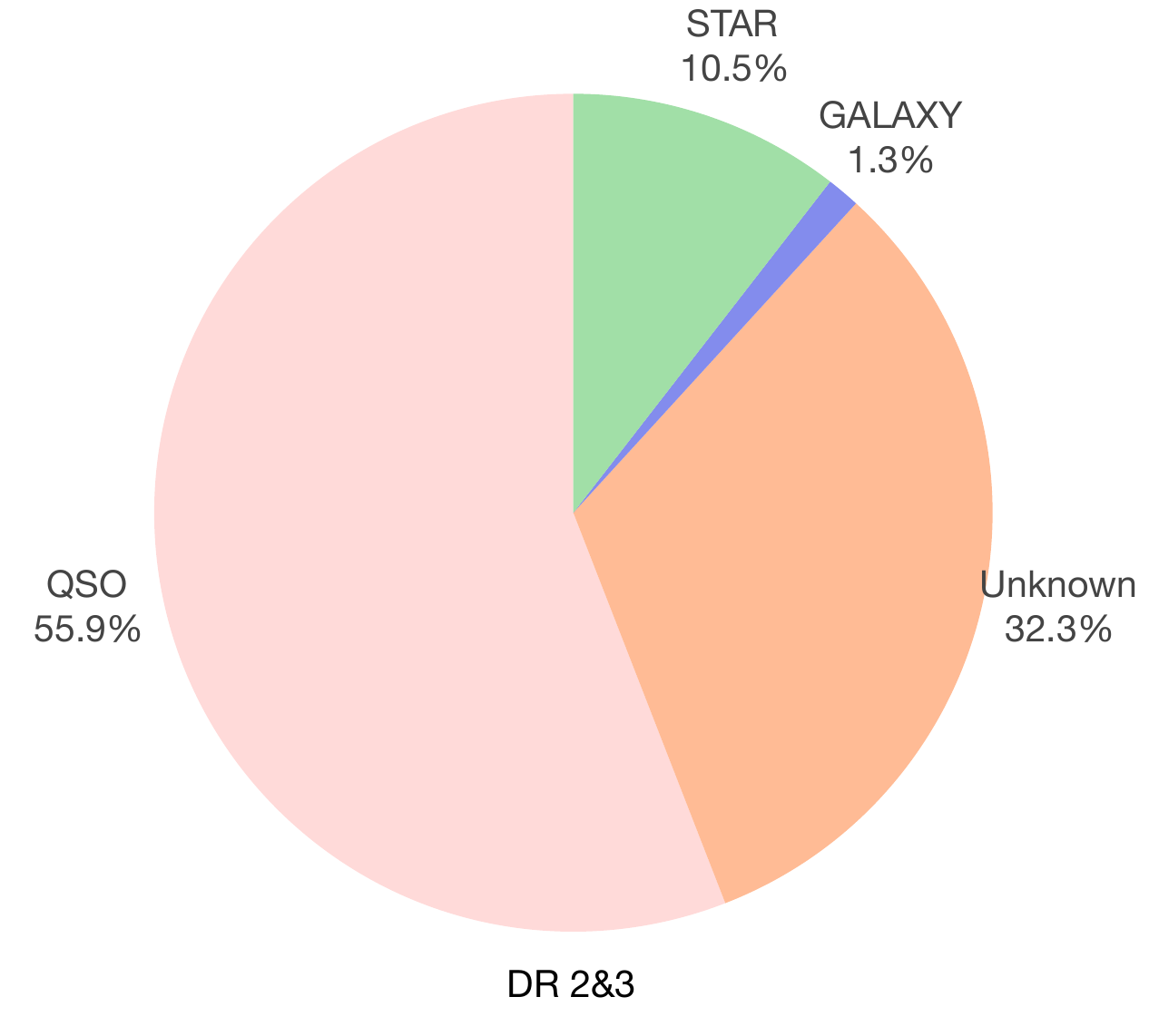} &
		\includegraphics[width=0.46\textwidth]{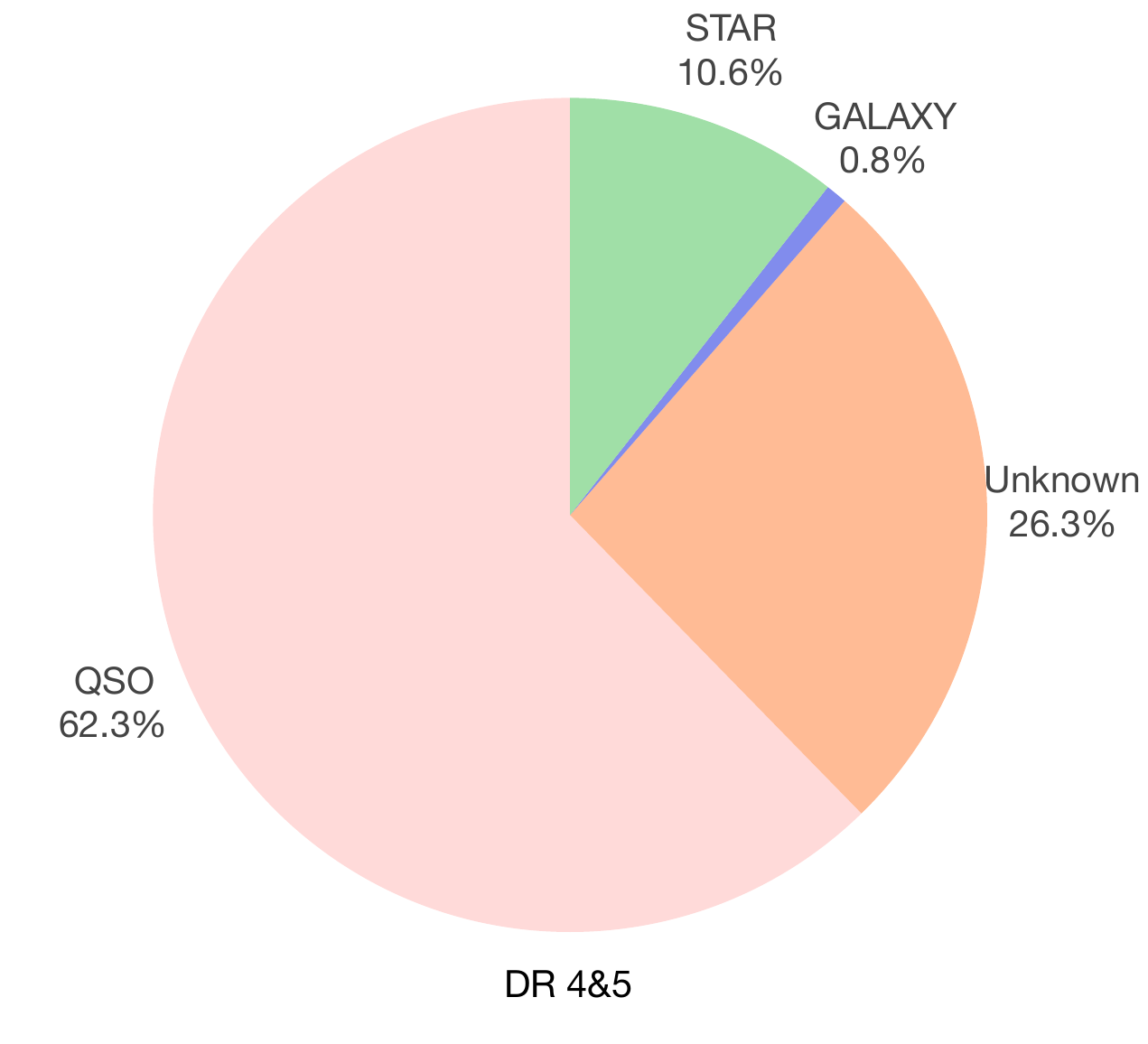} 
	\end{tabular}
	\caption{
	Fraction of objects classified as QSO, STAR, GALAXY and Unknown type by the pipeline among the targeted quasar candidates in the LAMOST regular survey. 
	}
\label{frac_type}
\end{figure*}

In the early data release, 
only $\sim14\%$ of the observed quasar candidates of the input catalog were identified as QSO, STAR or GALAXY by pipeline, 
while the majority of the spectra were classified as `Unknown' (Paper I). 
One of the main reason was that the magnitude limit of $i=20$ is too faint for the LAMOST LEGAS survey. 
The varying seeing 
due to the site conditions on one hand and the non-classical dome of the telescope on the other hand have significantly affected the spectrum quality \citep[][]{2012RAA....12..772Y, 2015RAA....15.1095L}. 
As a result, the identification was often limited by the poor quality of the spectrum. 
This can be clearly seen in DR1 (Paper I), 
where the pipeline `Unknowns' were on average one magnitude fainter than the pipeline identified ones. 
The observation conditions in later years have been improved since the first year regular survey. 
In addition, 
the pipeline has been updated in identifications of the spectra. 
The pie plots in Figure~\ref{frac_type} show the fractions of each type classified by pipeline among the observed quasar candidates. 
It is apparent that the fraction of `QSO' gets higher and higher in later data releases than in the early data release (Paper I). 

\begin{figure}
	\centering
	\includegraphics[width=\columnwidth]{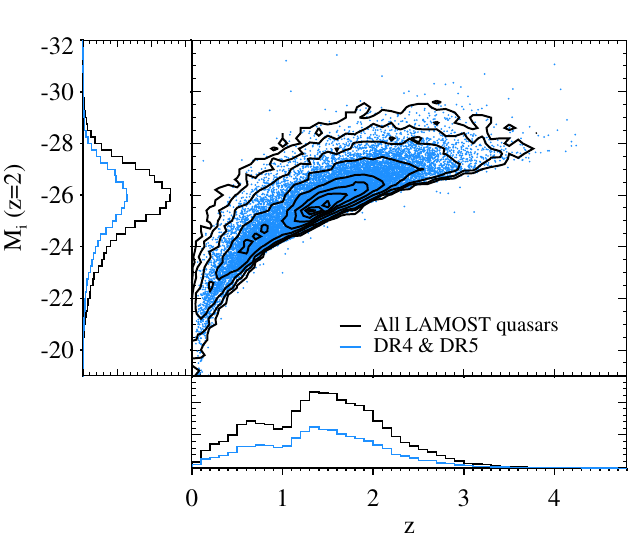}
	\caption{
	The absolute magnitude and redshift distribution for the visually confirmed quasars in whole LAMOST quasar survey (black) and in DR4 and DR5 of this paper (blue). 
	The absolute magnitudes $M_{i}(z=2)$ are normalized at $z=2$. 
	\label{z_lum}}
\end{figure}

We use a Java program {\tt ASERA} \citep[][]{2013A&C.....3...65Y} to help visually inspect the observed spectra of quasar candidates as well as the spectra that are classified as QSO by the 1D pipeline but not in the input catalog of quasar candidates. 
The misclassified spectra are rejected or re-classified by eye. 
Each spectrum is inspected by two or three classifiers to check if the features of spectrum can match the quasar template. 
We exclude the objects with only narrow emission lines by visual inspection. 
Finally, the objects identified as a quasar by at least two of the classifiers were included in the final quasar catalog. 
The redshift of each spectrum is also inspected and estimated according to available typical quasar emission lines. 
We estimate the redshift based on the peak of one of the emission lines 
in priority order of \oiiiwu, Mg~{\sc ii}, C~{\sc iii}, C~{\sc iv}, H$\beta$ and H$\alpha$-[N\,{\sc ii}], 
when any of them are available. 
A flag of `{\tt ZWARNING=1}' will be given when there is only one emission line visible. 
Finally, 
there are {19,253} quasars with reliable identifications in DR4 and DR5. 
Therefore, combining the results in previous data releases \citep[][]{2016AJ....151...24A, 2018AJ....155..189D}, 
the total number of identified quasars during the first five-year regular LAMOST quasar survey reaches {43,102}. 
LAMOST has also performed observations to search for the background quasars in the fields of M31 and M33, 
which are published elsewhere 
(e.g. \citealt{2010RAA....10..612H, 2013AJ....145..159H, 2015RAA....15.1438H}, and Huo et al. 2018 in preparation) 
and will not be included in this paper. 
Figure~\ref{z_lum} shows our identified quasars in the luminosity-redshift space, 
where the luminosity is indicated using the $K$-corrected $i$-band absolute magnitudes $M_{i}(z=2)$, normalized at $z=2$ \citep[][]{2006AJ....131.2766R}. 
As can be seen, there is a drop in the redshift distribution at $z\sim1.0$ for DR4 and DR5, similar to the results of previous LAMOST quasar data releases (Paper I and II). 
The drop comes from the inefficient identification of quasars in this redshift range when the Mg~{\sc ii} emission line moves around the overlapping wavelength region ($\sim6000$\,\AA) between the blue and red channels. 
The LAMOST quasar candidates were required to be point sources in our target selection (Section~\ref{selection}). 
It has been found that a significant fraction of quasars can be resolved up to $z\sim0.6$ in SDSS \citep[e.g.][]{2014ApJ...780..162M}. 
Thus, LAMOST quasar sample may suffer from incompleteness at low redshift $\lesssim0.6$. 

\begin{figure}
	\centering
	\includegraphics[width=\columnwidth]{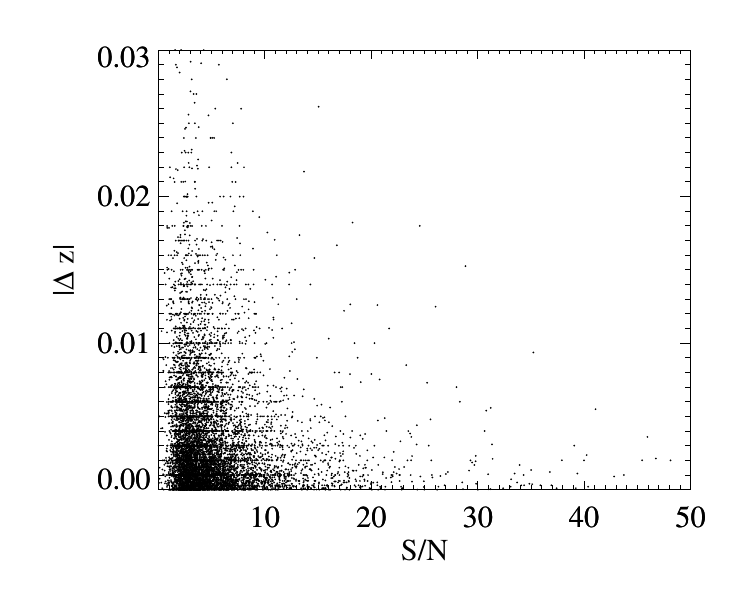}
	\caption{
	Redshift differences between LAMOST and SDSS as a function of LAMOST spectrum S/N. 
	\label{dz_snr}}
\end{figure}
Among the {19,253} identified quasars in DR4 and DR5, {11,097} are known ones in SDSS DR14 quasar catalog \citep{2018A&A...613A..51P} or NED\footnote{http://ned.ipac.caltech.edu/}, 
while the rest {8162} are newly discovered quasars by LAMOST. 
For the sources both observed in SDSS and LAMOST, 
only {119} of them have redshift differences $|\Delta z|=|z_{\rm LAMOST}-z_{\rm SDSS}|>0.1$. 
We have visually checked the spectra of these sources and find that the differences rise from the misidentification of the redshift for LAMOST spectra. 
For the other objects, 
the uncertainty of redshift may rise for two reasons. 
Firstly, we estimate the redshift based on one of the typical quasar emission lines in priority orders, 
while SDSS redshift is based on the result of a principal component analysis and the peak of the \mgii\ emission line \citep[][]{2018A&A...613A..51P}. 
In this case, 
sources of uncertainty could be line shifts relative to one another or line asymmetries. 
These cases are rare and deserve individual investigations. 
Secondly, the low S/N spectra also affect the estimation of redshift. 
In Figure~\ref{dz_snr}, 
we plot the redshift differences between LAMOST and SDSS as a function of LAMOST spectrum S/N. 
A increasing trend of $|\Delta z|$ with decreasing S/N can be clearly seen, 
indicating that low S/N could be a reason for the redshift uncertainties.

\begin{figure*}[t]
	\centering
	\includegraphics[width=0.98\textwidth]{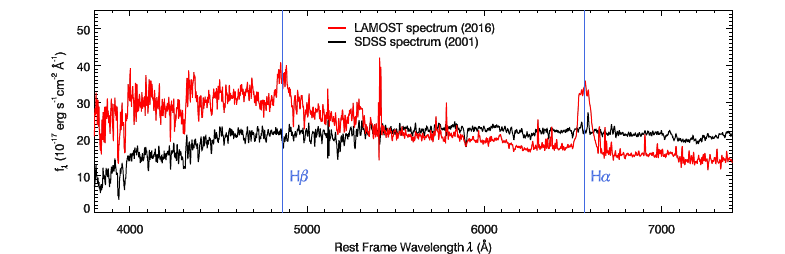}
	\caption{
	An example of the changing-look AGNs, J111536.57+054449.7 at $z=0.0897$ \citep{2018ApJ...862..109Y}, discovered by cross-matching the LAMOST and SDSS observations, which shows the emergence of its broad Balmer emission lines (labeled by blue vertical line). 
	The spectrum is corrected for the Galactic extinction and converted to the rest frame. 
	Here, the LAMOST spectrum is not corrected for the absolute flux calibration. 
	\label{clagn}}
\end{figure*}

As a large number of quasars in SDSS have also been observed by LAMOST, 
with time intervals of months to decades between SDSS and LAMOST survey, 
it is suitable to use SDSS and LAMOST observations to study the quasar spectral variability on both short and long timescales. 
For instance, it provides a large database for searching changing-look AGNs which shows appearance or disappearance of their broad emission lines \citep[e.g.,][]{2016A&A...593L...8M, 2016MNRAS.457..389M}. 
Actually, 
LAMOST has already produced many new discoveries of this type of rare objects \citep{2018ApJ...862..109Y}. 
In Figure~\ref{clagn} we present one such example of changing-look AGNs discovered by comparing LAMOST and SDSS observations. 

\section{Spectral Analysis} \label{sec:fitting}

In this and following sections we describe the spectral analysis, the measurements of typical quasar emission lines and the estimations of black hole masses. 
We note that LAMOST is only equipped with spectrograph and can not provide the photometric information for the observed targets. 
Thus, only relative flux calibration has been applied to the released LAMOST spectra. 
In addition, 
it is difficult to find a suitable flux standard star for each spectrograph during the observations of our extra-galactic targets 
since they are faint and located at high Galactic latitude. 
So the flux calibration is very poor for these targets due to the low S/N. 
Therefore, 
we use the SDSS photometric data to estimate the continuum luminosity (see Section~\ref{sec:bhmass}). 

We fit the spectra based on IDL routines in {\tt MPFIT} package \citep[][]{2009ASPC..411..251M}, 
which performs $\chi^{2}$ minimization using Levenberg-Marquardt method. 
Before the fitting, 
each spectrum is corrected for Galactic extinction using the reddening map of \citet{1998ApJ...500..525S} with an extinction curve of $R_V=3.1$ \citep{1999PASP..111...63F}, 
and then transformed into the rest frame of the object using the visually inspected redshift. 

Firstly, 
a broken power law is fitted to the spectrum in the wavelength windows of 
[1350, 1365]\,\AA, 
[1450, 1465]\,\AA, 
[1690, 1700]\,\AA, 
[3790, 3800]\,\AA, 
[4200, 4210]\,\AA, 
[5080, 5100]\,\AA, 
[5600, 5610]\,\AA, 
[6120, 6130]\,\AA~
and 
[6900, 6910]\,\AA. 
These windows are primarily chosen to avoid strong quasar emission lines listed in 
\citet[]{2001AJ....122..549V}. 
Pixels in the overlap region of LAMOST spectra (i.e. between 5700 to 6000]\,\AA~in the observed frame) between the red and blue channels are masked out during the fitting. 
The break of the power law is set to the value of 4661\,\AA\ at rest frame, 
which is similar to the value derived from the mean composite quasar spectra in \citet[]{2001AJ....122..549V}. 
The normalization and the indices are set as free parameters. 
In the next step, 
a local pseudo-continuum, i.e., a simple power law or a simple power law plus an Fe\,{\sc ii} template, 
and line models are used to separately fit each emission line we are interested in. 
Here the best-fit normalization and indices from the first step are taken as the initial guess values of power law normalization and index during the fitting in the second step. 

We are primarily interested in the broad H$\alpha$, H$\beta$, Mg\,{\sc ii} and C\,{\sc iv} emission lines, 
mainly because they are calibrated as virial black hole mass estimators and are the strongest broad emission lines in the available spectral range for most of the sources. 
The fitting procedures for each line are similar as those in Paper I 
(see also \citealt{2008MNRAS.383..581D} and \citealt{2009ApJ...707.1334W}) 
and described as follows. 

\subsection{H$\alpha$ and H$\beta$}

We fit the H$\alpha$ emission line for objects at $z\lesssim0.33$ and H$\beta$ emission line for the quasars at $z\lesssim0.76$. 
The profiles and redshifts of \hb\ and \ha\ are not tied together since they are fitted separately. 
In order to fit H$\beta$, 
a pseudo-continuum consisting of a simple power law plus Fe\,{\sc ii} multiplet is used to fit 
the spectrum in the windows [4435, 4700]\,\AA~and [5100, 5535]\,\AA. 
The Fe\,{\sc ii} template from \citet{2004A&A...417..515V} is adopted to model the optical Fe\,{\sc ii} multiplets. 
Then Gaussian profiles are used to model H$\beta$+[O\,{\sc iii}] emission lines in the range of [4600, 5100]\,\AA\ after subtracting the best-fit pseudo-continuum from the spectrum. 
We use two Gaussians to account for the broad H$\beta$ component and one Gaussian profile to account for the narrow H$\beta$ component. 
The upper limit of the full width at half maximum (FWHM) for the narrow component is set to be $900\rm\,km\,s^{-1}$. 
It was shown in previous studies that some AGNs reveal blue wings and blueshifts in [O\,{\sc iii}] \citep[e.g.,][]{2005ApJ...627..721G, 2007ApJ...667L..33K}. 
So each line of the [O\,{\sc iii}]$\lambda\lambda$4959,5007 doublet is modeled by two Gaussians, 
one for a line core and the other for a blueshifted wing component, 
and they are not tied to the \hb\ narrow component. 
The profile and redshift of each line of [O\,{\sc iii}]$\lambda\lambda$4959,5007 doublet is tied during the fitting and their flux ratio is fixed at the theoretical value of 1:3.

\begin{figure}[t]
	\centering
	\includegraphics[width=0.48\textwidth]{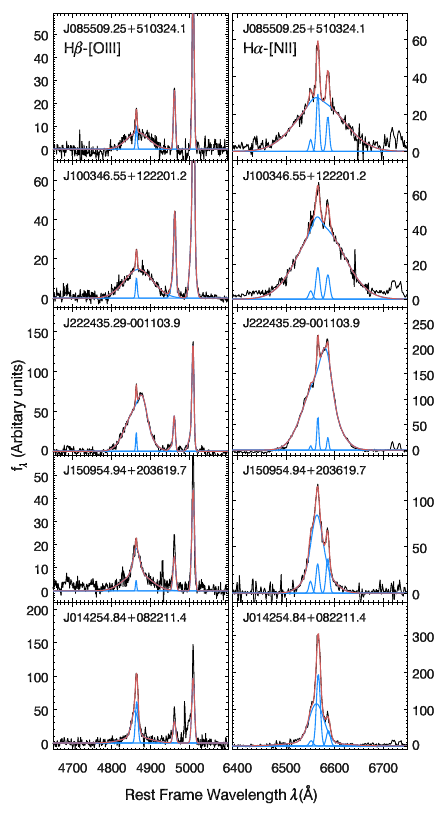}
	\caption{
	Sample results of the fitting procedures applied to H$\beta$+[O\,{\sc iii}] ({\it left panel}) and H$\alpha$+[N\,{\sc ii}] ({\it right panel}) lines. 
	The black lines represent the extinction-corrected spectra after subtracting the best-fit pseudo-continuum. 
	The blue lines represent the emission line model and the red lines represent the combination of models. 
	\label{hahb}}
\end{figure}

For the H$\alpha$, 
a same pseudo-continuum is applied to fit the spectrum in windows of [6000, 6250]\,\AA~and [6800, 7000]\,\AA. 
The normalization and the index of the power law are set to be free, while the parameters of the Fe\,{\sc ii} template are fixed at the best-fit values obtained from the H$\beta$ fitting. 
We model the pseudo-continuum-subtracted H$\alpha$-[N\,{\sc ii}]-[S\,{\sc ii}] emission lines in the spectral range of [6350, 6800]\,\AA\ using Gaussian profiles. 
The broad component of H$\alpha$ is modeled by two Gaussians, 
whereas [N\,{\sc ii}]$\lambda\lambda$6548,6584, [S\,{\sc ii}]$\lambda\lambda$6716,6731 and narrow H$\alpha$ component are modeled by a single Gaussian profile. 
The upper limits of the FWHM for the narrow lines are set to be $900\rm\,km\,s^{-1}$. 
The profile and redshift of the narrow lines are tied to each other 
and the relative flux ratio of [N\,{\sc ii}]$\lambda\lambda$6548,6584 doublet is fixed at 2.96. 
Examples of the best-fitting results are given in Figure~\ref{hahb}. 

\subsection{Mg\,{\sc ii}}

Fitting of the Mg\,{\sc ii} and C\,{\sc iv} emission lines are sometimes affected by the absorption features. 
Similar to Paper I, 
in order to reduce the effect of narrow absorption features, we mask out 3$\sigma$ outliers below the 20 pixel boxcar-smoothed spectrum when fitting Mg\,{\sc ii} and C\,{\sc iv} emission lines. 

\begin{figure}[t]
	\centering
	\includegraphics[width=0.48\textwidth]{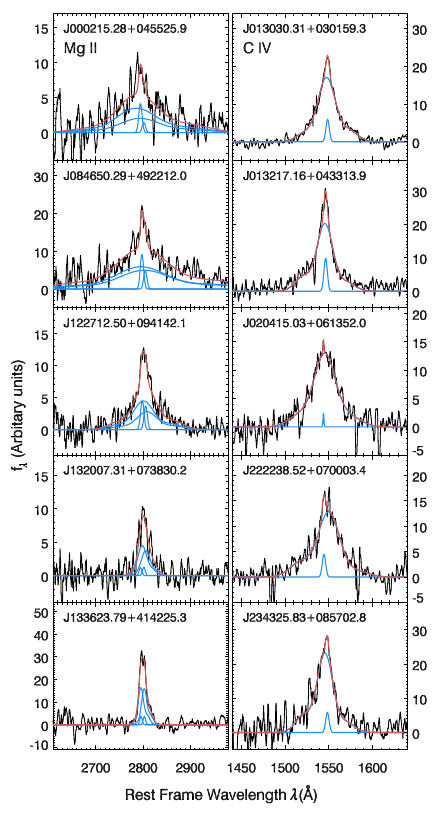}
	\caption{
	Same as in Figure~\ref{hahb} but for Mg\,{\sc ii} doublet ({\it left panel}) and C\,{\sc iv} ({\it right panel}) lines. 
	\label{mgii_civ}}
\end{figure}

We fit Mg\,{\sc ii} emission line for objects at $0.4\lesssim z\lesssim2.0$. 
A pseudo-continuum consisting of a simple power law and ultraviolet Fe\,{\sc ii} multiplet are used to fit the spectrum in the range of [2500, 2700]\,\AA~and [2900, 3090]\,\AA. 
The Fe\,{\sc ii} multiplet beneath the Mg\,{\sc ii} is modeled with semi-empirical template generated by \citet{2006ApJ...650...57T}. 
The best-fit pseudo-continuum is subtracted from the spectrum. 
Then we fit the Mg\,{\sc ii}$\lambda\lambda$2796,2803 doublet using a similar model as in \citet{2009ApJ...707.1334W}. 
Each of the narrow components is modeled with a single Gaussian profile. 
The upper limit of FWHM is set to be $900\rm\,km\,s^{-1}$. 
The width and redshift of the narrow lines are tied to each other. 
The flux ratio is constrained between 2:1 and 1:1 \citep[][]{1997ApJ...489..656L}. 
Each of the broad components is modeled by a truncated five-parameter Gauss-Hermite series (\citealt{1993ApJ...407..525V}; see also \citealt{2007ApJ...662..131S}). 
The profile, redshift and flux ratio of two broad components are constrained following the same prescription as for the narrow components. 
Examples of the best-fitting results are given in left panel of Figure~\ref{mgii_civ}.

\subsection{C\,{\sc iv}}

We fit the C\,{\sc iv} emission line for objects at $1.8\lesssim z\lesssim4.3$. 
Only a simple power law is fitted to the spectrum in range of [1445, 1465]\,\AA~and [1690, 1705]\,\AA~as the pseudo-continuum 
because Fe\,{\sc ii} multiplet underneath C\,{\sc iv} is very weak and 
it was found that fitting C\,{\sc iv} without subtraction of Fe\,{\sc ii} does not change the FWHM of C\,{\sc iv} significantly \citep[][hereafter S11]{2011ApJS..194...45S}. 
After subtracting the best-fit continuum from the spectrum, 
a Gauss-Hermite series and single Gaussian profile are used to model the broad and narrow components, respectively. 
Similar to the fitting procedure in Paper I, 
we do not set upper limit for FWHM of the narrow component as there are still debates on whether or not a strong narrow component exists for the C\,{\sc iv} line \citep{2005MNRAS.356.1029B}. 
Examples of the best-fitting results are given in the right panel of Figure~\ref{mgii_civ}. 

\subsection{Reliability of the Fittings and Error Estimation}
\label{sec:reliability}

After the automatic procedure, 
we inspect the fitting results visually for each object, 
pick out the bad fittings and fine-tuning the fitting procedures. 
Finally, the results are acceptable for most of the spectra with high S/N. 
The bad fittings are mainly caused by low S/N or lack of good pixels. 
For each line, 
a flag of {\tt LINE\_FLAG=0} is given to indicate an acceptable fitting and reliable measurement, 
while {\tt LINE\_FLAG=-1} indicates a spurious measurement. 
We note that 
a emission line is not fitted when the fitting wavelength range is not fully covered by the observed wavelength range, 
or when there is not enough good pixels ($n_{\rm pixel}<100$) due to bad quality of the spectrum, 
or when the fitting wavelength range of the line is red-shifted to the overlapping region between the blue and red channels. 
In this case, 
we set {\tt LINE\_FLAG=-9999}.
The broad absorption features at Mg\,{\sc ii} and C\,{\sc iv} also affect some of the fitting results. 
We give a flag of {\tt BAL\_FLAG} to indicate whether broad absorption features are present in the spectra by visual inspection. 

\begin{figure}[t]
	\centering
	\includegraphics[width=0.48\textwidth]{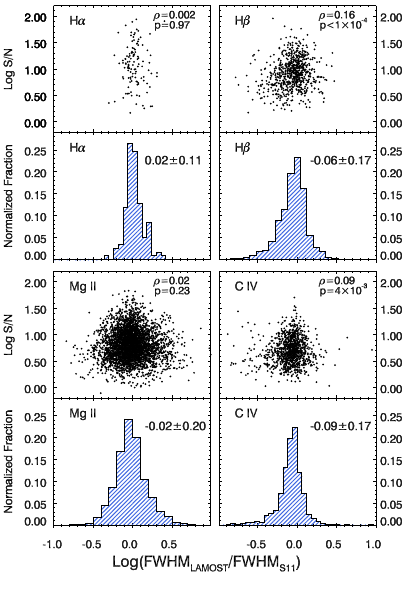}
	\caption{
	Comparison between the measurements of line width 
	in this work and in \citet{2011ApJS..194...45S}. 
	We show the plot of $\log(\rm FWHM_{LAMOST}/FWHM_{S11})$ versus the median S/N per pixel in line-fitting region 
	and the normalized distribution of $\log(\rm FWHM_{LAMOST}/FWHM_{S11})$ 
	for broad H$\alpha$ (upper left), broad H$\beta$ (upper right), broad Mg\,{\sc ii} (lower left) and total C\,{\sc iv} (lower right) emission lines, respectively. 
	The Spearman correlation coefficient $\rho$, $p$-values, mean value of the distribution and its standard deviation 
	are also tabulated in the corresponding plots. 
	Only the spectra with reliable fit ({\tt LINE\_FLAG=0}) are considered. 
	\label{fig:width_snr}}
\end{figure}

\begin{figure}[t]
	\centering
	\includegraphics[width=0.48\textwidth]{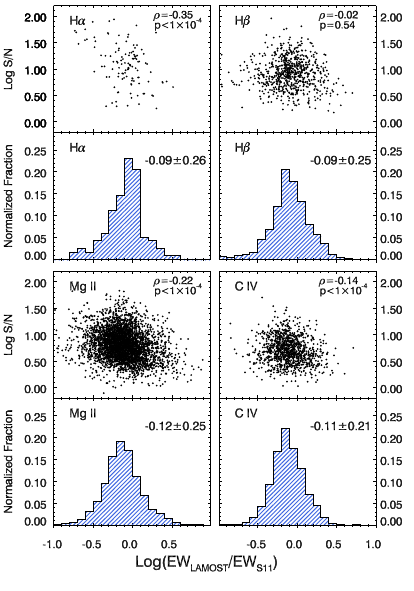}
	\caption{
	Same as Figure~\ref{fig:width_snr}, 
	but for comparison between the measurements of equivalent width 
	in this work and in \citet{2011ApJS..194...45S}. 
	\label{fig:ew_snr}}
\end{figure}

In Figure~\ref{fig:width_snr} and \ref{fig:ew_snr} we plot the comparison between our fitting results 
and those of S11 for 5974 overlapped sources. 
As can be seen from the normalized distribution of $\log(\rm LAMOST/S11)$, 
though being generally consistent with S11, 
our measurement values 
appear to be systematically smaller (see the mean value and standard deviation of the distribution in figures). 
As mentioned in Paper II, 
the differences may be caused by different models used in the fitting. 
We use double Gaussians to model the broad H$\alpha$ and H$\beta$ lines, 
and use the Gauss-Hermite series to model the broad component of Mg\,{\sc ii} and total C\,{\sc iv}, 
while in S11 multiple Gaussians with up to three Gaussians are used to fit each broad emission lines. 
The templates of iron multiplet are also different, 
which may lead to a different pseudo-continuum. 

\begin{figure}[t]
	\centering
	\includegraphics[width=0.48\textwidth]{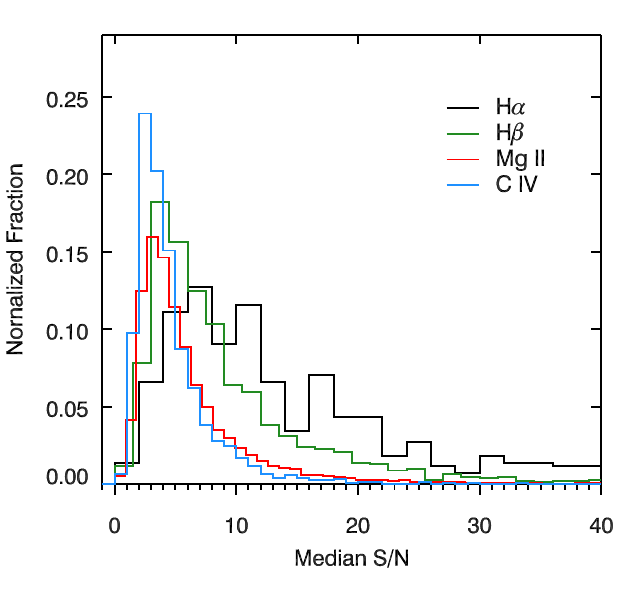}
	\caption{
	Normalized distributions of the median S/N per pixel around the line-fitting regions for objects with line measurements. 
	Only the spectra with reliable fit ({\tt LINE\_FLAG=0}) are considered. 
	\label{dist_linesnr}}
\end{figure}

\begin{figure}[t]
	\centering
	\includegraphics[width=0.48\textwidth]{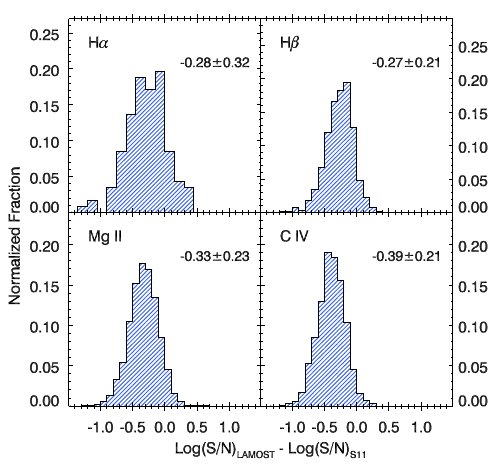}
	\caption{
	Comparison between the median S/N per pixel in the line-fitting region in this work and in \citet{2011ApJS..194...45S}. 
	The mean value and the standard deviation of distribution are tabulated in corresponding plots. 
	Only the spectra with reliable fit ({\tt LINE\_FLAG=0}) are considered. 
	\label{lm_shen_linesnr}}
\end{figure}

Another possible reason is the different S/N. 
Figure~\ref{dist_linesnr} shows the normalized distributions of the median S/N per pixel in line-fitting regions for quasars in LAMOST DR4\&5. 
The peak of the distributions 
are all around or below $\rm S/N=5$. 
We compare the median S/N per pixel of the line-fitting regions for the overlapped sources in LAMOST DR4\&5 and S11 in Figure~\ref{lm_shen_linesnr}. 
The normalized distributions of $\log(S/N)_{\rm LAMOST}-\log(S/N)_{\rm S11}$ show that 
our spectra have significant lower S/N than those in S11 for the same object. 
To explore the impact of S/N on our fitting results, 
we also show the plots of $\log(\rm LAMOST/S11)$ versus the median S/N per pixel in line-fitting regions. 
In Figure~\ref{fig:width_snr}, 
a trend of decreasing $\log(\rm FWHM_{LAMOST}/FWHM_{S11})$ with decreasing S/N is demonstrated by Spearman correlation test ($\rho>0$), 
although this trend is not significant for \ha\ and \mgii. 
In Figure~\ref{fig:ew_snr}, 
a trend of decreasing $\log(\rm EW_{LAMOST}/EW_{S11})$ with increasing S/N is demonstrated ($\rho<0$) for \ha\ \mgii\ and \civ. 
The simulation by S11 expects that 
the measured FWHMs and EWs are biased by less than $\pm20\%$ for high-EW objects 
as S/N decreases, 
while the FWHMs and EWs are biased low/high by $>20\%$ 
as S/N decreases. 
However, 
the exact dependences of FWHMs and EWs on spectrum S/N are probably more complicated and need further explorations in the future.

To estimate the errors of the measured FWHM and EW for the broad emission lines, 
we assume that the measured quantities of each component of the broad emission lines follow the Gaussian distribution, 
of which the standard deviation is the uncertainty obtained from $\chi^{2}$ minimization. 
We randomly generate each component and measure the FWHM and EW of profile combined by these components. 
The process was repeated 1000 times and the uncertainties of FWHM and EW for the broad emission line were estimated from the 68\% range of the distributions. 
For the narrow emission lines, the errors of the measurement is obtained from the $\chi^{2}$ minimization. 

\section{Continuum Luminosity and Virial Black Hole Mass} \label{sec:bhmass}

\begin{figure}[t]
	\centering
	\includegraphics[width=0.49\textwidth]{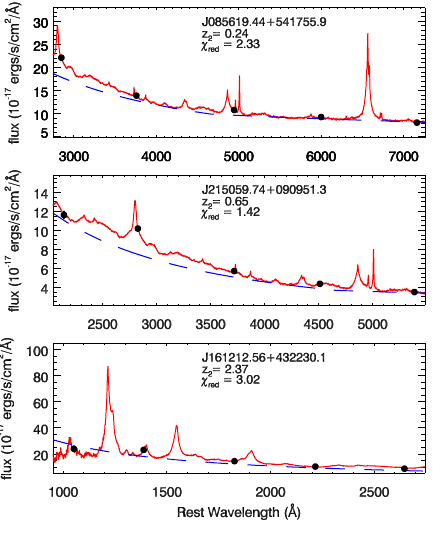}
	\caption{
	Examples of the model fitting to the SDSS $ugriz$-band photometry (black dots). 
	The red solid line represents the total model spectra and the blue dashed line represents the power laws used to estimated the continuum emission. 
	\label{pho_model}}
\end{figure}

In order to estimate the continuum luminosity for LAMOST quasars, 
we use the SDSS $ugriz$-band photometry following the procedures in Paper I and II. 
Firstly, 
we cross-match LAMOST DR4/DR5 quasars with SDSS photometric database with a 2\arcsec-matching-radius. 
We extract the PSF magnitudes, correct them for Galactic extinction, and convert them into the flux densities $F_{\lambda}$ at the effective wavelength of each filter. 
Next, we fit the five flux densities in the rest frame with a quasar spectrum model 
which is composed of a broken power-law continuum emission and line emission made from the composite quasar spectra in \citet{2001AJ....122..549V}. 
During the fitting, 
the break of the broken power laws is fixed at 4661\,\AA~and the normalization of the power laws is tied to the mean value of emission in ranges of [1350-1365]\,\AA~and [4200-4230]\,\AA~(Paper I). 
The indices of the power laws and the normalization of the line emission are set as free parameters. 
As results, the average power-law indices from the best-fits are found to be $\langle\alpha_{\nu}\rangle=-0.40$ ($\langle\alpha_{\lambda}\rangle=-1.60$) blueward of 4661\,\AA\ and $\langle\alpha_{\nu}\rangle=-1.65$ ($\langle\alpha_{\lambda}\rangle=-0.35$) redward of 4661\,\AA, which is in good agreement with those of the median composite spectra in \citet{2001AJ....122..549V}. 
Examples of the fitting results are presented in Figure~\ref{pho_model}. 

The monochromatic continuum luminosities at 1350\,\AA, 3000\,\AA~and 5100\,\AA~are calculated from the best-fit power-law continuum, which can be used as proxies for the radius of the broad emission line region. 
Meanwhile, with the broad line width as a proxy for the virial velocity, 
the black hole masses can be estimated based on the empirical scaling relations between the virial black hole mass, the FWHM of primary emission lines and the corresponding continuum luminosities, 
which are calibrated using local AGNs with reverberation mapping masses \citep[e.g.,][]{2004MNRAS.352.1390M, 2006ApJ...641..689V, 2009ApJ...707.1334W}. 
Here we estimate the virial black hole masses using relations
\begin{equation}
\begin{split}
	\log&M_{\rm BH}=\\
	\log&
	\left\{
	\left[ 
		\frac{\rm FWHM(H\beta)}{\rm km\,s^{-1}} 
	\right]^{2} 
	\left[
		\frac{L_{5100}}{10^{44}\rm\,ergs\,s^{-1}}
	\right]^{0.5} 
	\right\}
	+0.91
\end{split}
\end{equation}
for H$\beta$-based estimates \citep{2006ApJ...641..689V}, 
\begin{equation}
\begin{split}
	\log&M_{\rm BH}=\\
	\log&
	\left\{
	\left[ 
		\frac{\rm FWHM(Mg\,\text{\sc ii})}{\rm km\,s^{-1}} 
	\right]^{1.51} 
	\left[
		\frac{L_{3000}}{10^{44}\rm\,ergs\,s^{-1}}
	\right]^{0.5} 
	\right\}
	+2.60
\end{split}
\end{equation}
for Mg\,{\sc ii}-based estimates \citep{2009ApJ...707.1334W}, 
and 
\begin{equation}
\begin{split}
	\log&M_{\rm BH}=\\
	\log&
	\left\{
	\left[ 
		\frac{\rm FWHM(C\,\text{\sc iv})}{\rm km\,s^{-1}} 
	\right]^{2} 
	\left[
		\frac{L_{1350}}{10^{44}\rm\,ergs\,s^{-1}}
	\right]^{0.53} 
	\right\}
	+0.66
\end{split}
\end{equation}
for the C\,{\sc iv}-based estimates \citep{2006ApJ...641..689V}. 
The FWHM of broad H$\beta$, broad Mg\,{\sc ii} and whole C\,{\sc iv} line are used during the estimations. 

\begin{figure}[t]
	\centering
	\includegraphics[width=0.48\textwidth]{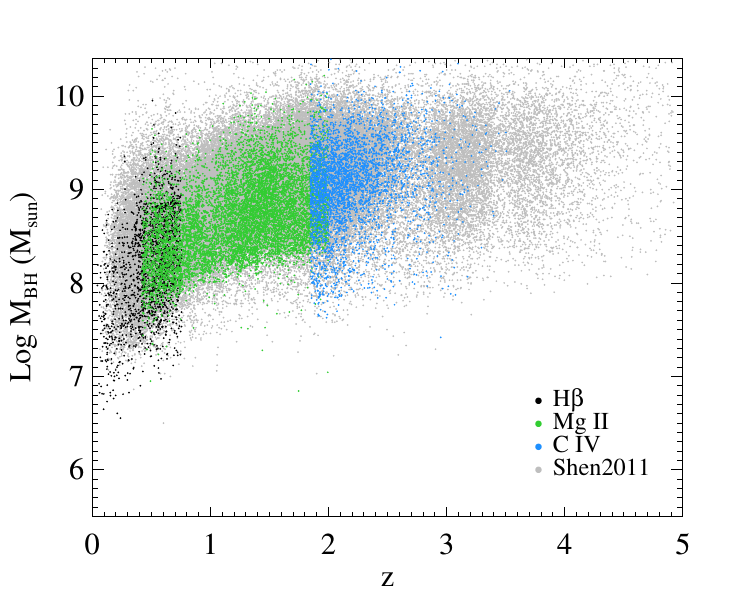}
	\caption{
	Black hole mass distribution along the redshift for the visually confirmed quasars of LAMOST DR4\&5 with estimated black hole masses based on H$\beta$ (black), Mg{\sc\,ii} (green) and C{\sc\,iv} (blue). 
	The grey dots are SDSS quasars from 
	\citet{2011ApJS..194...45S}. 
	\label{zmass}}
\end{figure}

\begin{figure}[t]
	\centering
	\includegraphics[width=0.48\textwidth]{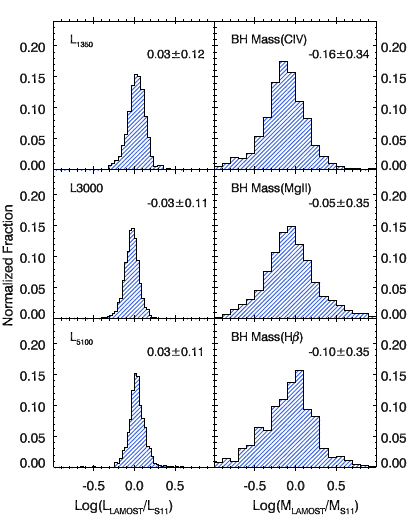}
	\caption{
	{\it Left:} Comparison (normalized distributions) of the continuum luminosities in this work and in \citet{2011ApJS..194...45S}. 
	{\it Right:} Comparison of the black hole masses estimated from H$\beta$, Mg\,{\sc ii} and C\,{\sc iv} emission lines in this work and in \citet{2011ApJS..194...45S}. 
	\label{lum_mass}}
\end{figure}

Although it's straightforward to calculate the black hole mass using the above relations, 
one should bear in mind the large uncertainties of the estimates ($\gtrsim0.4\rm\,dex$, e.g., \citealt{2006ApJ...641..689V, 2009ApJ...707.1334W}). 
Figure~\ref{zmass} shows the distribution of the black hole masses with redshifts. 
It is apparent that 
mass estimates for the LAMOST quasars occupy the same area as those of SDSS DR7 quasars from S11. 
We also note that our estimations adopt the continuum luminosities obtained from the SDSS photometry more than ten years before the LAMOST survey, 
which introduce additional uncertainties because the ultraviolet/optical emission of quasars varied generally with magnitudes of $0.1-0.2\rm\,mag$. 
To justify this effect, 
for the overlapped objects in LAMOST DR4\&5 and S11, 
we compare our estimates of continuum luminosities and black hole masses with those estimated by S11 in Figure~\ref{lum_mass}. 
It is shown that our estimates are in general agreement with S11's results, 
though with systematically lower values than S11. 
The lower value of black hole mass may be driven by the lower S/N leading to smaller line width, 
as mentioned in Section~\ref{sec:reliability}. 

\section{Description of the Catalog} \label{sec:catalog}

We give a compiled catalog for the quasars identified in LAMOST DR4\&5 along with this paper. 
The keyword for each column are listed in Table~\ref{tab:catalog} and described as below.

\begin{itemize}[noitemsep]
  \item[1.] Unique spectra ID. 
  \item[2.] Target Observation date. 
  \item[3.] LAMOST DR4\&5 object name: $\rm Jhhmmss.ss+ddmmss.s$ (J2000). 
  \item[4-5.] Right Ascension and Declination (in decimal degrees, J2000). 
  \item[6-9.] Information of the spectroscopic observation: 
                   Modified Julian date (MJD), spectroscopic plan name (PlanID), spectrograph identification (spID), and spectroscopic fiber number (fiberID). 
                   These four numbers are unique for each spectrum named in the format of 
                   {spec$-$MJD$-$planID\_spID$-$fiberID.fits}. 
  \item[10-11.] Redshift and its flag based on visual inspections. 1$=$not robust. 
  \item[12.] Target selection flag. 
  		 This flag is a four-character string indicating how the quasar candidate is selected, 
		 in which the value of each character is a boolean value, i.e., ``1'' is truth and ``0'' is false. 
		 The first character indicates infrared-optical color selection. 
		 The second character indicates data-mining  selection. 
		 The third character indicates selection based on X-ray detection. 
		 The fourth character indicates selection based on radio detection. 
		 A entry with SOURCE\_FLAG=``1101'' indicates that the quasar was selected by its infrared-optical color, data-mining algorithms and radio detection. 
		 A entry with SOURCE\_FLAG=``0000'' means the object is not included in LAMOST LEGAS quasar survey sample but identified as quasar. 
  \item[13.] $M_{i}(z=2)$: absolute i-band magnitude. K-corrected to $z=2$ following \citet{2006AJ....131.2766R}. 
  \item[14.] Number of spectroscopic observations for the quasar. When there are more than one observation for the object, the line properties are obtained from only one of the observations in which the fiber pointing position is nearest to the position in our quasar candidate catalog. 
  \item[15.] Median S/N per pixel in the wavelength regions of [4000-5700]\,\AA\ and[6200, 9000]\,\AA. 
  \item[16.] Flag of broad absorption features. BAL\_FLAG$=$1 indicates broad absorption features are present. 
  \item[17-30.] FWHM, rest-frame equivalent width, and their uncertainties, for broad \ha, narrow \ha, [N\,{\sc ii}]$\lambda6584$, [S\,{\sc ii}]$\lambda\lambda$6716,6731 emission lines. 
  \item[31.] Rest-frame equivalent width of iron emissions in 6000-6500\,\AA.
  \item[32-33.] Number of good pixels and median S/N per pixel for the spectrum in region of rest-frame 6400-6765\,\AA. 
  \item[34.] Flag indicates reliability of the emission line fitting results in \ha\ region upon visual inspections. 0$=$reliable; -1$=$unreliable. 
                  This value is set to be $-9999$ if \ha\ is not measured due to lack of good pixels in the fitting wavelength region. 
  \item[35-46.] FWHM, rest-frame equivalent width, and their uncertainties, for broad \hb, narrow \hb, [O\,{\sc iii}]$\lambda\lambda$5007 emission lines. 
  \item[47] Rest-frame equivalent width of iron emissions in 4685-4435\,\AA. 
  \item[48-49.] Number of good pixels and median S/N per pixel for the spectrum in region of rest-frame 4750-4950\,\AA. 
  \item[50.] Flag indicates reliability of the emission line fitting results in \hb\ region upon visual inspections. 0$=$reliable; -1$=$unreliable. 
                  This value is set to be $-9999$ if \hb\ is not measured due to lack of good pixels in the fitting wavelength region. 
  \item[51-54.] FWHM, rest-frame equivalent width, and their uncertainties, of the whole \mgii\ emission line. 
  \item[55-58.] FWHM, rest-frame equivalent width, and their uncertainties, of the total broad \mgii\ emission line. 
  \item[59-62.] FWHM and its uncertainties of the broad and narrow \mgii$\lambda$2796 emission lines. 
  \item[63.] Rest-frame equivalent width of iron emissions in 2200-3090\,\AA. 
  \item[64-65.] Number of good pixels and median S/N per pixel for the spectrum in region of rest-frame 2700-2900\,\AA. 
  \item[66.] Flag indicates reliability of the emission line fitting results in \mgii\ region upon visual inspections. 0$=$reliable; -1$=$unreliable. 
                  This value is set to be $-9999$ if \mgii\ is not measured due to lack of good pixels in the fitting wavelength region. 
  \item[67-70.] FWHM, rest-frame equivalent width, and their uncertainties, of the whole \civ\ emission line. 
  \item[71-74.] FWHM, rest-frame equivalent width, and their uncertainties, of the broad \civ\ emission line. 
  \item[75-78.] FWHM, rest-frame equivalent width, and their uncertainties, of the narrow \civ\ emission line. 
  \item[79-80.] Number of good pixels and median S/N per pixel for the spectrum in region of rest-frame 1500-1600\,\AA. 
  \item[81.] Flag indicates reliability of the emission line fitting results in \civ\ region upon visual inspections. 0$=$reliable; -1$=$unreliable. 
                  This value is set to be $-9999$ if \civ\ is not measured due to lack of good pixels in the fitting wavelength region.
  \item[82.] Wavelength power-law index, $\alpha_\lambda$, from $\sim$ 1300 to 4661\,\AA. 
  \item[83.] Wavelength power-law index, $\alpha_\lambda$, from $\sim$ redward of 4661\,\AA. 
  \item[84.] Reduced chi-square in SDSS photometry modeling; $-9999$ if not fitted. 
  \item[85-87.] Monochromatic luminosities at 5100\,\AA, 3000\,\AA\ and 1350\,\AA. 
  \item[88-90.] Virial black hole masses (in $M_{\odot}$) based on \hb, \mgii\ and \civ. 
  \item[91] Name of the quasar in SDSS quasar catalog. 
  The LAMOST DR4\&5 quasar catalog was cross-correlated with the SDSS quasar catalog \citep[DR14,][]{2018A&A...613A..51P} using a matching radius of 3\arcsec. 
  \item[92] Name of the object in 2nd ROSAT all-sky survey point source catalog \citep[2RXS,][]{2016A&A...588A.103B}. 
  The LAMOST DR4\&5 quasar catalog was cross-correlated with 2nd ROSAT all-sky survey point source catalog using a matching radius of 30\arcsec. 
  The nearest point source in 2RXS was chosen. 
  \item[93-94] The background corrected source counts in full band (0.1-2.4\,keV), and its error, from 2nd ROSAT all-sky survey point source catalog \citep[][]{2016A&A...588A.103B}. 
  \item[95] The exposure time of the ROSAT measurement. 
  \item[96] Angular separation between the LAMOST and ROSAT source positions. 
  \item[97] Name of the object in XMM-Newton Serendipitous Source Catalog \citep[3XMM-DR8,][]{2016A&A...590A...1R}. The LAMOST DR4\&5 quasar catalog was cross-correlated with XMM-Newton Serendipitous Source Catalog using a matching radius of 3\arcsec. 
  \item[98-99] The mean full-band (0.2-12\,keV) flux, and its error, from XMM-Newton Serendipitous Source Catalog \citep[][]{2016A&A...590A...1R}. 
  \item[100] Angular separation between the LAMOST and 3XMM-DR8 source positions. 
  \item[101] FIRST peak flux density at 20\,cm in units of mJy. 
  The LAMOST DR4\&5 quasar catalog was cross-correlated with FIRST survey catalog using a matching radius of 5\arcsec. 
  \item[102] Angular separation between LAMOST and FIRST source positions. 
\end{itemize}

\section{Summary} \label{sec:summary}

\begin{figure}[t]
	\centering
	\includegraphics[width=0.44\textwidth]{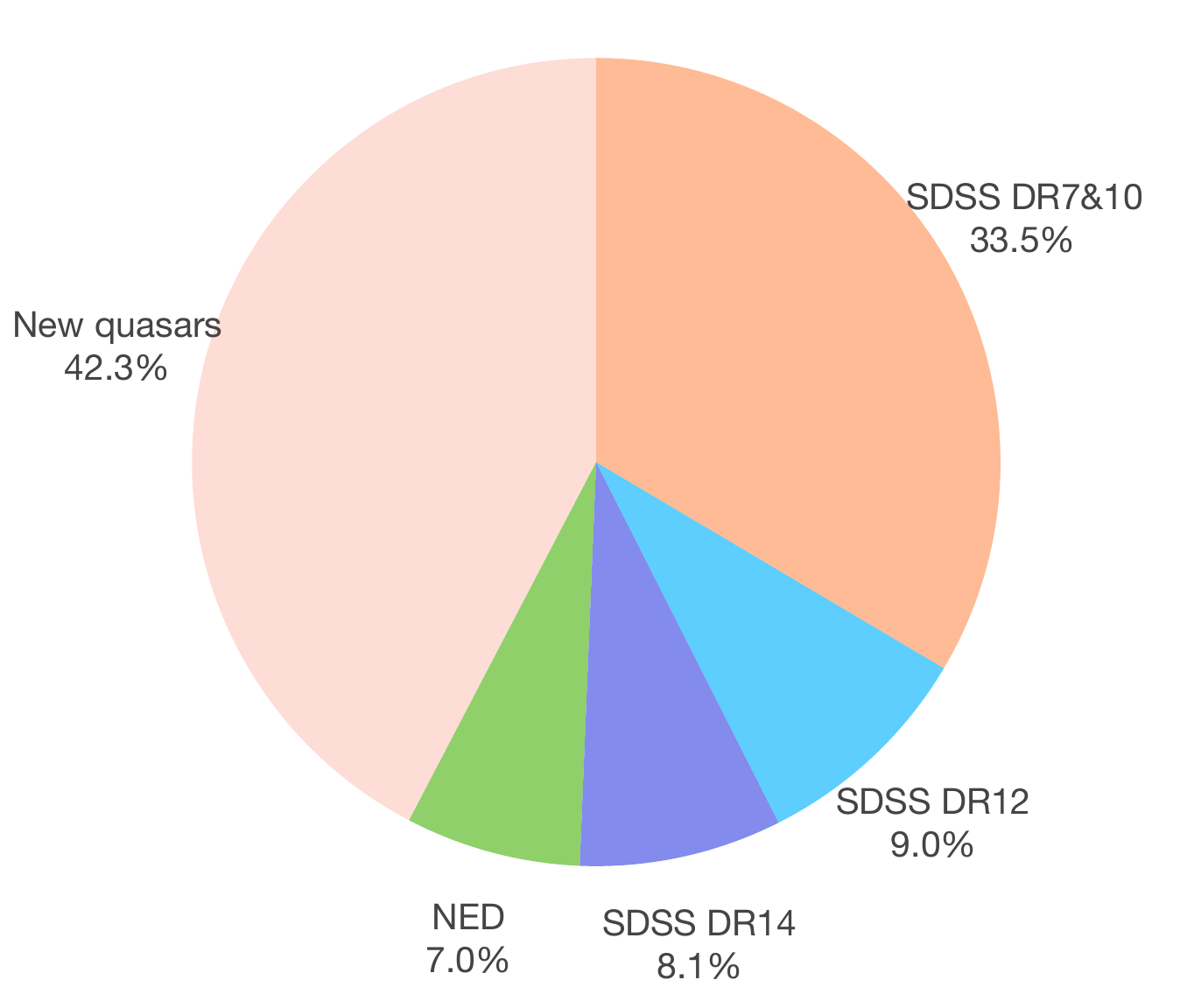}
	\caption{
	Visually confirmed quasars in LAMOST DR4\&5. 
	Nearly half of the sources are newly discovered. 
	\label{dr45}}
\end{figure}

In this paper, 
we present the results of the LAMOST quasar survey in the 4th and 5th data release obtained from the observations which began at September 2015 and September 2016, respectively. 
There are a total of {19,253} visually confirmed quasars. 
The catalog and spectra of these quasars will be available online.
Among these identified quasars, 
{11,458} were independently discovered by LAMOST, 
{3296} of which were reported in SDSS DR12/14 quasar catalog \citep[][]{2017A&A...597A..79P, 2018A&A...613A..51P} after the survey began, 
while the other {8162} are new discoveries of LAMOST that have not been reported before (see Figure~\ref{dr45}). 
After the first five-year regular observation, 
the total number of identified quasars in LAMOST quasar survey reaches {43,102} hitherto 
\citep[][and this work]{2016AJ....151...24A, 2018AJ....155..189D}. 
There are {24,772} independently discovered quasars by LAMOST, 
and {17,128} of them are newly discovered.

In this work, 
we provide \ha, \hb, \mgii\ and \civ\ emission line properties (FWHM and equivalent width) for each quasar spectrum by performing spectral analysis. 
Although LAMOST is not equipped with a photometry instrument and lack the information about absolute flux calibration for the spectra, 
we obtain the continuum luminosity underneath the emission lines 
and the corresponding black hole masses by fitting the SDSS photometric data using the quasar spectra.

In addition to the great supplement to the low-to-moderate redshift quasar discoveries, 
LAMOST also provides a large database for investigating the quasar spectral variabilities 
since the LAMOST and SDSS observations were taken over roughly ten-years baseline. 
By comparing the LAMOST and SDSS data, more unusual quasars with, 
including changing-look AGNs \citep[][]{2018ApJ...862..109Y} or tidal disruption events \citep[][]{2018arXiv180306362L}
may be discovered.


\acknowledgments

This work is supported by 
the National Key Basic Research Program of China 2014CB845700,
the Ministry of Science and Technology of China under grant 2016YFA0400703, 
the NSFC grants No.11721303 and 11533001. 
Su Yao acknowledges the support by the KIAA-CAS Fellowship, 
which is jointly supported by Peking University and Chinese Academy of Sciences, 
and the PKU Boya fellowship. 
Y. L. Ai acknowledges the support by grant No. U1731109. 
Guoshoujing Telescope (the Large Sky Area Multi-Object Fiber Spectroscopic Telescope LAMOST) is a National Major Scientific Project built by the Chinese Academy of Sciences. Funding for the project has been provided by the National Development and Reform Commission. LAMOST is operated and managed by the National Astronomical Observatories, Chinese Academy of Sciences.

This publication makes use of data products from the Sloan Digital Sky Survey. 
Funding for the Sloan Digital Sky Survey IV has been provided by the Alfred P. Sloan Foundation, the U.S. Department of Energy Office of Science, and the Participating Institutions. SDSS-IV acknowledges
support and resources from the Center for High-Performance Computing at
the University of Utah. The SDSS web site is www.sdss.org.

SDSS-IV is managed by the Astrophysical Research Consortium for the 
Participating Institutions of the SDSS Collaboration including the 
Brazilian Participation Group, the Carnegie Institution for Science, 
Carnegie Mellon University, the Chilean Participation Group, the French Participation Group, Harvard-Smithsonian Center for Astrophysics, 
Instituto de Astrof\'isica de Canarias, The Johns Hopkins University, 
Kavli Institute for the Physics and Mathematics of the Universe (IPMU) / 
University of Tokyo, Lawrence Berkeley National Laboratory, 
Leibniz Institut f\"ur Astrophysik Potsdam (AIP),  
Max-Planck-Institut f\"ur Astronomie (MPIA Heidelberg), 
Max-Planck-Institut f\"ur Astrophysik (MPA Garching), 
Max-Planck-Institut f\"ur Extraterrestrische Physik (MPE), 
National Astronomical Observatories of China, New Mexico State University, 
New York University, University of Notre Dame, 
Observat\'ario Nacional / MCTI, The Ohio State University, 
Pennsylvania State University, Shanghai Astronomical Observatory, 
United Kingdom Participation Group,
Universidad Nacional Aut\'onoma de M\'exico, University of Arizona, 
University of Colorado Boulder, University of Oxford, University of Portsmouth, 
University of Utah, University of Virginia, University of Washington, University of Wisconsin, 
Vanderbilt University, and Yale University.

%

\vspace{5mm}
\facilities{LAMOST}

\startlongtable
\begin{deluxetable*}{llll}
\tabletypesize{\scriptsize} 
\tablecaption{Catalog format for the quasars identified in LAMOST DR4\&5\label{tab:catalog}}
\tablehead{\colhead{Column} &  \colhead{Name} &  \colhead{Format} &  \colhead{Description}}  
\startdata
1   & ObsID  &  LONG  & Unique Spectra ID \\
2   & ObsDate & STRING  &  Target observation date \\
3   & NAME  &  STRING  &  LAMOST designation hhmmss.ss+ddmmss (J2000) \\
4   & RA            &  DOUBLE  &  Right ascension in decimal degrees (J2000) \\
5   & DEC           &  DOUBLE  &  Declination in decimal degrees (J2000) \\
6   & LMJD           &  LONG    &  Local Modified Julian Day of observation \\
7   & PLANID       &  STRING  &  Spectroscopic plan name \\
8   & SPID          &  LONG  &  Spectrograph identification \\
9   & FIBERID       &  LONG    &  Spectroscopic fiber number \\
\hline
10 & Z$\_$VI          &  DOUBLE  &  Redshift from visual inspection \\
11 & ZWARNING    &  LONG    &  ZWARNING flag from visual inspection \\
{12}	&  {SOURCE\_FLAG}  & STRING   &  Target selection flag \\
13 & MI$\_$Z2         &  DOUBLE  &  M$_{i} (z=2)$, K-corrected to $z=2$ following Richards et al. (2006) \\
{14}  & {NSPECOBS}	              & LONG  &  Number of spectroscopic observations \\
15  & SNR\_SPEC               &  DOUBLE &  Median S/N per pixel of the spectrum \\
16  &  BAL\_FLAG        & LONG  &  Flag of broad absorption features \\
\hline
17  & FWHM\_BROAD\_HA	      &  DOUBLE	&  FWHM of broad \ha\ in \kmps \\
{18}  & {ERR\_FWHM\_BROAD\_HA}    &  DOUBLE	&  Uncertainty in FWHM$_{\rm H\alpha,broad}$ \\
19	& EW\_BROAD\_HA	          &  DOUBLE	&  Rest-frame equivalent width of broad \ha\ in \AA \\
{20}	& {ERR\_EW\_BROAD\_HA}	&  DOUBLE	&  Uncertainty in EW$_{\rm H\alpha,broad}$ \\
21	& FWHM\_NARROW\_HA	      &  DOUBLE	&  FWHM of narrow \ha\ in \kmps \\
{22}	& {ERR\_FWHM\_NARROW\_HA}	  &  DOUBLE	&  Uncertainty in FWHM$_{\rm H\alpha,narrow}$ \\
23  & EW\_NARROW\_HA	      &  DOUBLE	&  Rest-frame equivalent width of narrow \ha\ in \AA \\
{24}	& {ERR\_EW\_NARROW\_HA}	  &  DOUBLE	&  Uncertainty in EW$_{\rm H\alpha,narrow}$ \\
25	& EW\_NII\_6584           &  DOUBLE	&  Rest-frame equivalent width of [N\,{\sc ii}]$\lambda$6584 in \AA \\
{26}	& {ERR\_EW\_NII\_6584}	  &  DOUBLE	&  Uncertainty in EW$_{\rm[NII]6584}$ \\
27	& EW\_SII\_6716        &  DOUBLE	&  Rest-frame equivalent width of [S\,{\sc ii}]$\lambda$6716 in \AA \\
{28}	& {ERR\_EW\_SII\_6718}	  &  DOUBLE	&  Uncertainty in EW$_{\rm[SII]6716}$ \\
29	& EW\_SII\_6731         &  DOUBLE	&  Rest-frame equivalent width of [S\,{\sc ii}]$\lambda$6731 in \AA  \\
{30}	& {ERR\_EW\_SII\_6731}	  &  DOUBLE	&  Uncertainty in EW$_{\rm[SII]6731}$ \\
31	& EW\_FE\_HA	          &  DOUBLE	&  Rest-frame equivalent width of Fe within 6000-6500\,\AA\ in \AA \\
32	& LINE\_NPIX\_HA	      &  LONG	&  Number of good pixels for the rest-frame 6400-6765\,\AA  \\
33	& LINE\_MED\_SN\_HA	      &  DOUBLE	&  Median S/N per pixel for the rest-frame 6400-6765\,\AA   \\
34  & LINE\_FLAG\_HA          &  LONG   &  Flag for the quality in \ha\ fitting \\
\hline
35  & FWHM\_BROAD\_HB	      &  DOUBLE	&  FWHM of broad \hb\ in \kmps \\
{36}  & {ERR\_FWHM\_BROAD\_HB}    &  DOUBLE	&  Uncertainty in FWHM$_{\rm H\beta,broad}$ \\
37	& EW\_BROAD\_HB	          &  DOUBLE	&  Rest-frame equivalent width of broad \hb\ in \AA  \\
{38}	& {ERR\_EW\_BROAD\_HB}	  &  DOUBLE	&  Uncertainty in EW$_{\rm H\beta,broad}$ \\
39	& FWHM\_NARROW\_HB	      &  DOUBLE	&  FWHM of narrow \hb\ in \kmps \\
{40}	& {ERR\_FWHM\_NARROW\_HB}	  &  DOUBLE	&  Uncertainty in FWHM$_{\rm H\beta,narrow}$ \\
41  & EW\_NARROW\_HB	      &  DOUBLE	&  Rest-frame equivalent width of narrow \hb\ in \AA \\
{42}	& {ERR\_EW\_NARROW\_HB}	  &  DOUBLE	&  Uncertainty in EW$_{\rm H\beta,narrow}$ \\
{43}	& {FWHM\_OIII\_5007}   &  DOUBLE	 &  FWHM of [O\,{\sc iii}]$\lambda$5007 in \kmps \\
{44}	& {ERR\_FWHM\_OIII\_5007}  & DOUBLE & Uncertainty in FWHM$_{\rm[OIII]5007}$ \\
45	& EW\_OIII\_5007	      &  DOUBLE	&  Rest-frame equivalent width of [O\,{\sc iii}]$\lambda$5007 in \AA \\
{46}	& {ERR\_EW\_OIII\_5007}     &	 DOUBLE	&  Uncertainty in EW$_{\rm[OIII]5007}$ \\
47	& EW\_FE\_HB              &  DOUBLE	&  Rest-frame equivalent width of Fe within 4435-4685\,\AA\ in \AA \\
48	& LINE\_NPIX\_HB	      &  LONG	&  Number of good pixels for the rest-frame 4750-4950\,\AA  \\
49	& LINE\_MED\_SN\_HB	      &  DOUBLE	&  Median S/N per pixel for the res t-frame 4750-4950\,\AA  \\
50  & LINE\_FLAG\_HB          &  LONG   &  Flag for the quality in \hb\ fitting\\
\hline
51	& FWHM\_MGII              &  DOUBLE	&  FWHM of the whole \mgii\ emission line in \kmps \\
{52}	& {ERR\_FWHM\_MGII}      &  DOUBLE	&  Uncertainty in FWHM$_{\rm MgII,whole}$ \\
53	& EW\_MGII                &	 DOUBLE	&  Rest-frame equivalent width of the whole \mgii\ in \AA \\
{54}	& {ERR\_EW\_MGII}	   &  DOUBLE	&  Uncertainty in EW$_{\rm MgII,whole}$ \\
55	& FWHM\_BROAD\_MGII	      &  DOUBLE	&  FWHM of the whole broad \mgii\ in \kmps \\
{56}	& {ERR\_FWHM\_BROAD\_MGII}  &  DOUBLE	&  Uncertainty in FWHM$_{\rm MgII,broad}$  \\
57	& EW\_BROAD\_MGII	      &  DOUBLE	&  Rest-frame equivalent width of the whole broad \mgii\ in \AA \\
{58}	& {ERR\_EW\_BROAD\_MGII}	  &  DOUBLE	&  Uncertainty in EW$_{\rm MgII,broad}$ \\
59  & FWHM\_BROAD\_MGII\_2796 &  DOUBLE &  FWHM of the broad \mgii$\lambda$2796 in \kmps \\
{60}  & {ERR\_FWHM\_BROAD\_MGII\_2796} &  DOUBLE &  Uncertainty in FWHM$_{\rm MgII2796,broad}$ \\
61  & FWHM\_NARROW\_MGII\_2796      &  DOUBLE &  FWHM of the narrow \mgii$\lambda$2796 in \kmps \\
{62}  & {ERR\_FWHM\_NARROW\_MGII\_2796} &  DOUBLE &  Uncertainty in FWHM$_{\rm MgII2796,narrow}$ \\
63  & EW\_FE\_MGII	          &  DOUBLE	&  Rest frame equivalent width of Fe within 2200-3090\,\AA\ in \AA \\
64	& LINE\_NPIX\_MGII        &	 LONG	&  Number of good pixels for the rest-frame 2700-2900\,\AA  \\
65	& LINE\_MED\_SN\_MGII	  &  DOUBLE	&  Median S/N per pixel for the rest-frame 2700-2900\,\AA   \\
66  & LINE\_FLAG\_MGII        &  LONG   &  Flag for the quality in MgII fitting \\
\hline
67	& FWHM\_CIV	              &  DOUBLE &	FWHM of the whole \civ\ in \kmps \\
{68}	& {ERR\_FWHM\_CIV}    &  DOUBLE	&   Uncertainty in FWHM$_{\rm CIV,whole}$ \\
69	& EW\_CIV	              &  DOUBLE	&   Rest-frame equivalent width of the whole \civ\ in \AA \\
{70}	& {ERR\_EW\_CIV}	   &  DOUBLE	&   Uncertainty in EW$_{\rm CIV,whole}$ \\
71	& FWHM\_BROAD\_CIV	      &  DOUBLE &	FWHM of the broad \civ\ in \kmps \\
{72}	& {ERR\_FWHM\_BROAD\_CIV}  &  DOUBLE	&   Uncertainty in FWHM$_{\rm CIV,broad}$ \\
73	& EW\_BROAD\_CIV	      &  DOUBLE	&   Rest-frame equivalent width of the broad \civ\ in \AA \\
{74}	& {ERR\_EW\_BROAD\_CIV}	  &  DOUBLE	&   Uncertainty in EW$_{\rm CIV,broad}$ \\
75	& FWHM\_NARROW\_CIV	      &  DOUBLE &	FWHM of the narrow \civ\ in \kmps \\
{76}	& {ERR\_FWHM\_NARROW\_CIV}  &  DOUBLE	&   Uncertainty in FWHM$_{\rm CIV,narrow}$ \\
77	& EW\_NARROW\_CIV	      &  DOUBLE	&   Rest-frame equivalent width of the narrow \civ\ in \AA \\
{78}	& {ERR\_EW\_NARROW\_CIV}	  &  DOUBLE	&   Uncertainty in EW$_{\rm CIV,narrow}$ \\
79  & LINE\_NPIX\_CIV	      &  LONG	&   Number of good pixels for the rest-frame 1500-1600\,\AA  \\
80	& LINE\_MED\_SN\_CIV	  &  DOUBLE	&   Median S/N per pixel for the rest-frame 1500-1600\,\AA   \\
81  & LINE\_FLAG\_CIV         &  LONG   &  Flag for the quality in CIV fitting \\
\hline
82    & ALPHA\_LAMBDA\_1   &  DOUBLE &   Wavelength power-law index from $\sim$ 1300 to 4661\,\AA \\
83    & ALPHA\_LAMBDA\_2   &  DOUBLE &   Wavelength power-law index redward of 4661\,\AA \\
84    & MODEL\_PHOT\_REDCHI2 &  DOUBLE  & Reduced chi-square  \\ 
85	& LOGL5100	              & DOUBLE	&   Monochromatic luminosity at 5100\,\AA\ in $\rm erg\,s^{-1}$ \\
86	& LOGL3000	              & DOUBLE	&   Monochromatic luminosity at 3000\,\AA\ in $\rm erg\,s^{-1}$ \\
87	& LOGL1350	              & DOUBLE	&   Monochromatic luminosity at 1350\,\AA\ in $\rm erg\,s^{-1}$ \\
88	& LOGBH\_HB            &	DOUBLE	&   Virial BH mass (M$_{\sun}$) based on \hb \\
89	& LOGBH\_MgII          &	DOUBLE	&   Virial BH mass (M$_{\sun}$) based on \mgii \\
90	& LOGBH\_CIV           &	DOUBLE	&   Virial BH mass (M$_{\sun}$) based on \civ \\
\hline
{91}	& {SDSS\_NAME}	& STRING & Name of the quasar in the SDSS quasar catalog (P{\^a}ris et al. 2018) \\
{92}	& {2RXS\_NAME}	& STRING &  Name of the object in the 2nd ROSAT all-sky survey point source catalog (2RXS, Boller et al. 2016) \\
{93}	& {2RXS\_CTS} 	& DOUBLE & Background corrected source counts in 0.1-2.4\,keV from 2RXS source catalog \\
{94}	& {2RXS\_ECTS} 	& DOUBLE & Error of the source counts from 2RXS source catalog \\
{95}	& {2RXS\_EXPTIME} & DOUBLE & Source exposure time from 2RXS source catalog \\
{96}	& {LM\_2RXS\_SEP} & DOUBLE & LAMOST-2RXS separation in arcsec \\
{97}	& {3XMM\_NAME} & STRING & Name of the object in XMM-Newton Serendipitous Source Catalog (3XMM-DR8, Rosen et al. 2016) \\
{98}	& {3XMM\_FLUX} 	& DOUBLE & Flux in 0.2-12.0\,keV band from 3XMM-DR8 (in erg\,s$^{-1}$\,cm$^{-2}$) \\
{99}	& {3XMM\_FLUX\_ERR} 	& DOUBLE & Error of the flux in 0.2-12.0\,keV band from 3XMM-DR8 (in erg\,s$^{-1}$\,cm$^{-2}$) \\
{100}	& {LM\_3XMM\_SEP} & DOUBLE & LAMOST-3XMM separation in arcsec \\
{101}	& {FPEAK}	& DOUBLE & FIRST peak flux density at 20 cm in mJy \\
{102}	& {LM\_FIRST\_SEP}	& DOUBLE & LAMOST-FIRST separation in arcsec \\
\enddata
\tablenotetext{}{{(This table is available in its entirety in FITS format.)}}
\end{deluxetable*}

\bibliographystyle{aasjournal}
\bibliography{references}



\end{document}